\documentclass[prl,twocolumn,showpacs,superscriptaddress]{revtex4}
\usepackage{graphicx,bm,amssymb,amsmath}

\newcommand{\Eq}[1]{Eq.~\eqref{#1}}
\newcommand{\eq}[1]{\eqref{#1}}

\newcommand{\pdag}{{\phantom{\dagger}}}

\newcommand{\past}{{\phantom{\ast}}}

\newcommand{\beq}{\begin{equation}}
\newcommand{\eeq}{\end{equation}}

\newcommand{\PRB}[1]{Phys. Rev. B~\textbf{#1}}
\newcommand{\PRL}[1]{Phys. Rev. Lett.~\textbf{#1}}
\newcommand{\RMP}[1]{Rev. Mod. Phys.~\textbf{#1}}

\newcommand{\etal}{\textit{et al.}} 

\begin{document}

\title{Phase transition, spin-charge separation, and spin filtering in a quantum dot}

\author{Michael Pustilnik}
\affiliation{School of Physics, Georgia Institute of Technology, 
Atlanta, GA 30332}

\author{L\'aszl\'o Borda}
\affiliation{Research Group ``Theory of Condensed Matter'' of the Hungarian
Academy of Sciences, Institute of Physics, TU Budapest,
Budafoki {\'u}t 8., H-1521, Hungary
}

\begin{abstract}
We consider low temperature transport through a lateral quantum 
dot asymmetrically coupled to two conducting leads, and tuned 
to the mixed-valence region separating two adjacent Coulomb 
blockade valleys with spin $S=1/2$ and $S=1$ on the dot. 
We demonstrate that this system exhibits a quantum phase transition
driven by the gate voltage. In the vicinity of the transition the spin 
on the dot is quantized, even though the fluctuations of charge are 
strong.  The spin-charge separation leads to an unusual Fano-like 
dependence of the conductance on the gate voltage and to an almost 
perfect spin polarization of the current through the dot in the presence 
of a magnetic field.
\end{abstract}

\pacs{
72.15.Qm,    
73.23.Hk,     
73.63.Kv      
}
\maketitle

In a single-electron transistor setup~\cite{blockade} the 
number of electrons $N$ in a quantum dot is controlled 
by the potential on the capacitively coupled gate 
electrode. At low temperature $N$ is close to an integer at almost 
any gate voltage, except narrow \textit{mixed-valence} regions, 
where adding a single electron to the dot does not lead to a 
large penalty in electrostatic energy. 
The distance between these regions sets the scale 
for the dependence of measurable quantities on the gate voltage, 
which makes it convenient to use a dimensionless parameter
$N_0$, the gate voltage normalized by this scale.
In terms of $N_0$,  the mixed-valence regions are narrow
intervals of the width
\beq
\Delta_N\sim\Gamma/E_C\ll 1,
\label{1}
\eeq
about half-integer values of $N_0$~\cite{review}.  
Here $E_C$ is the charging energy and $\Gamma$ is the 
tunneling-induced width of single-particle energy levels in the dot. 

In a typical experiment a dot is connected via tunneling junctions 
to two massive electrodes~\cite{blockade}. At temperatures in the 
range $\Gamma\lesssim T\ll E_C$, the conductance $G$ is suppressed 
outside the mixed-valence regions, resulting in a quasiperiodic 
sequence of well-defined Coulomb blockade peaks in the 
dependence $G(N_0)$~\cite{blockade,review}. When $T$ is further 
lowered, $G(N_0)$ changes dramatically due to the onset of the Kondo 
effect~\cite{review,unitary}. At $T\to 0$ pairs of adjacent Coulomb 
blockade peaks merge to form broad maxima at $N\!\approx\text{odd integer}$, 
separated by smooth crossovers from the minima at 
$N\!\approx\text{even integer}$. Although $G(N_0)$ at $T\gtrsim\Gamma$ 
is very different from that at $T\ll\Gamma$,  in the mixed-valence regions 
both functions are featureless~\cite{unitary,review}. 

The evolution of $G$ towards its low-temperature limit can 
be rather complicated. Indeed, GaAs quantum dots with odd 
$N$ usually have spin $S=1/2$~\cite{spin_1,spin_exp}. In this 
case the dependence $G(T)$ is characterized by a single energy 
scale, the Kondo temperature $T_K$; $G(T)$ increases 
monotonically with the decrease of $T$ at $T\ll\Gamma$~\cite{review}. 
However, dots with even $N$ often have spin $S=1$ rather 
than zero~\cite{spin_1,spin_exp,spin_theory}.  Kondo effect 
then occurs in two stages, controlled by two different energy scales, 
$T_K$ and $T_K^\prime<T_K$~\cite{real}. The resulting $G(T)$ 
is not monotonic: $G$ first raises, and then drops again when $T$ 
is lowered~\cite{real,two-stage}. The dependence of the conductance 
on the Zeeman energy $B$ of an applied magnetic field 
is also non-monotonic and is characterized by the same scales 
$T_K$ and $T_K^\prime$~\cite{real}.

The values of $T_K$ and $T_K^\prime$ depend on $N_0$ and
their ratio, in general, is not tunable. 
A notable exception occurs when the conductances 
of the dot-lead junctions are very different, i.e. when the width 
$\Gamma=\Gamma_L\!+\Gamma_R$ is dominated by the 
contribution from the lead with the stronger coupling to the dot, 
say, $\Gamma_L\gg\Gamma_R$. (Note that conductances of the 
junctions are easily tunable in lateral quantum dot systems 
such as those studied in~\cite{two-stage}). It can be shown~\cite{review} 
that in this limit $T_K^\prime\ll T_K$ for all $N_0$. In particular, in 
the vicinity of the mixed-valence region $T_K^\prime\sim\Gamma_R$ 
while $T_K\sim \Gamma_L\gg T_K^\prime$. Accordingly, the second 
stage of the Kondo effect will not develop if 
\beq
\Gamma_R\ll \max\{T,B\} \ll\Gamma_L.
\label{2}
\eeq 
In this paper we show that under these conditions the conductance
in the mixed-valence region between the Coulomb blockade valleys 
with $S=1/2$ and $S=1$ on the dot varies with $N_0$ on the scale 
which is parametrically small compared with $\Delta_N$, in striking
difference with the conventional smooth dependence described above.

The dependence of the conductance on $B$ at $B\gg T$ is 
qualitatively similar to its dependence on $T$ at $T\gg B$~\cite{review}. 
Since $B$-dependence is much easier to understand, we concentrate 
here on the limit $T\to 0$ (the effect of a finite $T$ is briefly 
discussed towards the end of the paper). 

The first inequality in \Eq{2} allows one to take into account the coupling 
to the right lead in the lowest non-vanishing order in $\Gamma_R/\Gamma_L$. 
The conductance at any finite $B$ is then given by~\cite{review}
\beq
g=g_\uparrow + g_\downarrow, 
\quad 
g_s = \sin^2\delta_s.
\label{3}
\eeq
Here $g=G/G_0$ is the conductance normalized by 
$G_0\sim (e^2/h)\Gamma_R/\Gamma_L$, the largest value 
conductance per spin can reach for strongly asymmetric 
coupling to the leads; $g_s$ is the conductance (in units of $G_0$) 
for electrons with spin $s$, and $\delta_s$ is scattering phase 
shift at the Fermi level for electrons with spin $s$ in the left lead. 

In order to calculate the phase shifts, we model a quantum 
dot coupled to a single lead by the Hamiltonian
\beq
H= H_0 + H_t + H_d.
\label{4}
\eeq
The first term here describes the electrons in the lead. 
For a lateral quantum dot it is sufficient to take into account 
only a single propagating mode~\cite{review,real},  
\beq
H_0 = \sum_{ks}\xi^\pdag_k \psi^\dagger_{ks} \psi^\pdag_{ks}, 
\label{5}
\eeq 
and the spectra $\xi_k$ can be linearized near the Fermi level, which 
corresponds to a constant density of states $\nu$.

The second term in \Eq{4} describes the tunneling coupling between 
the dot and the lead,
\beq
H_t = \sum_{nks}t_n^\pdag\psi^\dagger_{ks} d^\pdag_{ns} + {\rm H.c.}
\label{6}
\eeq
In the following we set $t_n\!=t$, so that all levels in the dot have the same 
width $\Gamma=\pi\nu t^2$. This assumption is not essential for the validity 
of the following consideration.
 
The last term in \Eq{4} describes an isolated dot,
\beq
H_d = \sum_{ns} \epsilon_n^\pdag d^\dagger_{ns} d^\pdag_{ns}
+ \,E_C(\hat N - N_0)^2 - E_S\hat{\bf S}^2 - B\hat S_z. 
\label{7}
\eeq
Here $\hat N =\sum_{ns} d^\dagger_{ns} d^\pdag_{ns}$
and 
$\hat{\bf S} 
= \frac{1}{2}\sum_{nss'} 
d^\dagger_{ns} 
\hat{\bm{\sigma}}_{ss'}
d^\pdag_{ns'}$ 
are operators of the total number of electrons on the dot, and of the 
dot's spin, respectively 
($\hat{\bm{\sigma}} = (\hat\sigma_x,\hat\sigma_y,\hat\sigma_z)$ 
are Pauli matrices). For a typical dot the parameters $\delta E$ 
(mean single-particle level spacing), $E_S$ (exchange energy), and 
$E_C$ (charging energy) satisfy $E_S\ll\delta E\ll E_C$~\cite{review}.

An isolated dot with even $N$ will have $S=1$ in the ground state if 
the spacing $\varepsilon$ between the two single-particle levels closest 
to the Fermi level is anomalously small, 
$2E_S-\varepsilon>0$~\cite{spin_1,spin_exp,spin_theory}.
For simplicity, we assume here that $2E_S-\varepsilon\gtrsim\Gamma$.
Although this simplification imposes a stronger restriction on $\Gamma$ 
then $\Gamma\ll\delta E$ (this inequality justifies the tunneling 
Hamiltonian description of the dot-lead junction~\cite{review}), 
it does not affect the results.

For the model \eq{4}-\eq{7} the phase shifts are given by
\beq
\delta_\uparrow = (\pi/2)(N+M),
\quad
\delta_\downarrow = (\pi/2)(N-M),
\label{8}
\eeq
where $N=\langle\hat N\rangle$ is the number of electrons 
in the dot, and $M=2\langle\hat S_z\rangle$ is the dot's
magnetization~\cite{phase_shifts}. 
We start with $N_0$ outside the mixed-valence region, 
\beq
\Delta_N\ll|N_0-\widetilde{N}_0|\ll 1,
\quad
\widetilde N_0 -1/2 =\text{odd integer},
\label{9}
\eeq
Here $N\approx \widetilde{N}_0\pm 1/2$ is close to an integer.
The tunneling-induced virtual transitions to states with ``wrong'' $N$ 
can be ``integrated out'' with the help of the Schrieffer-Wolff 
transformation, yielding an effective Kondo Hamiltonian
\beq
H = H_0 + V\!\rho + J\!\left({\bf s}\cdot\!{\bf S}\right) -BS_z,
\label{10}
\eeq
where $\rho=\sum_{kk's}\psi^\dagger_{ks} \psi^\pdag_{k's}$
and ${\bf s} = \frac{1}{2}\sum_{kk'ss'} 
\psi^\dagger_{ks} 
\hat{\bm{\sigma}}_{ss'}
\psi^\pdag_{k's'}$
are operators describing the local particle and spin densities 
of conduction electrons. The operator $\bf S$ in \Eq{10} 
is a projection of $\hat{\bf S}$ [see \Eq{7}] on the ground state 
multiplet of an isolated dot with fixed integer $N$.
The reduction of the microscopic model \eq{4}-\eq{7} to
the Kondo Hamiltonian \eq{10} is valid only when $N_0$ 
is outside the mixed-valence region and at sufficiently low energies, 
\beq
|\xi_k|\lesssim D=\min\Bigl\{d_N, \,2E_C\bigl|N_0-\widetilde{N}_0\bigr|\Bigr\},
\label{11}
\eeq
where $d_{N}=\delta E$ $(d_{N}=2E_C-\varepsilon)$ for odd (even) $N$.
The parameters $V$ and $J$ in \Eq{10} can be estimated as~\cite{review,real}
\beq
\nu V\!\sim \Delta_N\bigl(N_0-\widetilde{N}_0\bigr)^{-1},
\quad
\nu J\!\sim|\nu V| .
\label{12}
\eeq
(It should be noted that $V$ and $J$ are subject to strong mesoscopic 
fluctuations; the order-of-magnitude estimate \eq{12} is sufficient for 
our purpose).

The potential scattering term in \Eq{10} is responsible for the deviations 
$\delta N$ of the dot's occupation from the corresponding integer values 
$\widetilde{N}_0\pm 1/2$,
\beq
 \delta N = N-\bigl(\widetilde{N}_0\pm 1/2\bigr)
\approx - 2\nu V
\sim \Delta_N\bigl(\widetilde{N}_0-N_0\bigr)^{-1}.
\label{13}
\eeq
Note that $|\delta N|$ is finite and increases with approach 
to the mixed-valence region. Also note that a weak magnetic 
field $B\ll\Gamma$ does not affect $N(N_0)$. 

On the contrary, $M(N_0)$ depends strongly on $B$.
Indeed, $M(B)$ for a given $N_0$ is controlled by the 
Kondo temperature $T_K(N_0)$, which can be estimated from 
\beq
\ln\bigl(D/T_K\bigr)\sim(\nu J)^{-1}\!
\sim \Delta_N^{-1}\bigl|N_0-\widetilde{N}_0\bigr|.
\label{14}
\eeq
Accordingly, $T_K\sim\Gamma$ at 
$\bigl|N_0-\widetilde{N}_0\bigr|\sim\Delta_N$, and decreases 
exponentially with the increase of the distance to $\widetilde{N}_0$.
A fixed field $B$ is large compared to $T_K$ at 
$\bigl|N_0-\widetilde{N}_0\bigr|\gg \Delta_B$, where $\Delta_B$ 
is the distance between $N_0$ and $\widetilde{N}_0$
at which $B\sim T_K(N_0)$. In this regime
\beq
M/M_0 = 1 - \Bigl[2\ln(B/T_K)\Bigr]^{-1}, 
\label{15}
\eeq
where $M_0=1(2)$ for $N_0<\widetilde{N}_0$ $(N_0>\widetilde{N}_0)$.
In the opposite limit $\Delta_B\gg\bigl|N_0-\widetilde{N}_0\bigr|\gg\Delta_N$
(note that $\Delta_B\gg\Delta_N$ for $B\ll\Gamma$) the system is in 
the strong coupling regime $B\ll T_K$. Here $M(N_0)$ depends strongly 
on the parity of $N$. 
Indeed, $\bf S$ in \Eq{10} is spin-$1/2$ 
operator for odd $N$ (i.e. for $\widetilde{N}_0-N_0\gg\Delta_N$),
and spin-$1$ operator for even $N$ ($N_0-\widetilde{N}_0\gg\Delta_N$). 
This difference is crucial. An antiferromagnetic local exchange
interaction with a single species of itinerant electrons suffices 
to completely screen $S=1/2$ magnetic impurity, thereby 
forming a \textit{singlet} (non-degenerate) ground 
state~\cite{PWA,Nozieres,Bethe,Wilson}. 
In this case the approach to the low-energy fixed point
is Fermi-liquid-like~\cite{Nozieres}, 
and
\beq
M\sim 
B/T_K.
\label{16}
\eeq
On the contrary, for $S=1$ only half of the impurity's spin 
is screened, and the ground state is a \textit{doublet}~\cite{Nozieres,Bethe}. 
The low-energy physics is then described by the ferromagnetic 
exchange of the conduction electrons with the remaining spin 
$S=1/2$~\cite{Nozieres,Bethe}, 
and 
\beq
M = 1 + \Bigl[2\ln(T_K/B)\Bigr]^{-1}.
\label{17}
\eeq

\begin{figure}[h]
\includegraphics[width=0.8\columnwidth]{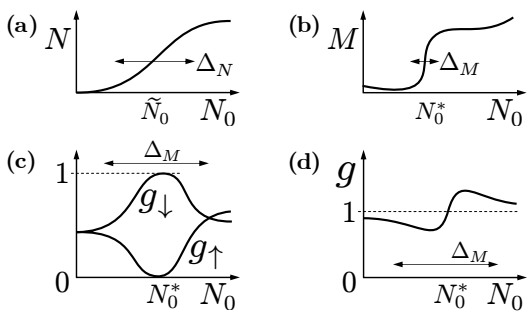}
\caption{
(a) Number of electrons in the dot $N$ differs appreciably from 
an integer in a narrow \textit{mixed-valence} region of the width 
$\Delta_N$. 
(b) At $B\ll\Gamma$, the width $\Delta_M$ of the \textit{crossover} 
region in the dependence of the magnetization $M$ on the gate voltage $N_0$ 
is small compared to $\Delta_N$. 
(c) Spin-resolved conductances in the crossover region at $T\ll B\ll\Gamma$.
(d) The total conductance at $\max\{B,T\}\ll\Gamma$.
}
\label{fig1}
\end{figure}

Although the above results were obtained for $N_0$ outside 
the mixed-valence region $\bigl|N_0-\widetilde{N}_0\bigr|\lesssim\Delta_N$,
some conclusions regarding this region can be drown as well. Indeed,
since $\delta N$ in \Eq{13} is finite, it is plausible that $N$ varies 
continuously with $N_0$, as sketched in Fig.~\ref{fig1}(a).  

The dependence $M(N_0)$ is more complicated. Consider the limit 
$B\to +0$. In this limit $M$ is determined solely by the ground state 
degeneracy. Since the degeneracy can not change continuously, 
the system must go through a \textit{quantum phase transition} (QPT) 
at a certain value of $N_0^\past=N_0^\ast$. As shown above, the ground 
state is either a singlet or a doublet when the charge fluctuations are weak.
Therefore, the transition must occur within the mixed-valence region,
i.e. $\bigl|N_0^\ast-\widetilde{N}_0^\past\bigr|\lesssim\Delta_N$.
The QPT manifests itself in a singular dependence of the magnetization 
$M$ on the gate voltage, 
\beq
\lim_{B\to+0} M = \theta(N_0^\past-N_0^\ast).
\label{18}
\eeq
Note that in the vicinity of the transition the spin is quantized, 
even though the fluctuations of charge are very strong, 
$N^*\!=N(N_0^\ast)\!\approx\text{half-integer}$ (unlike in the 
case of transitions that occur at a fixed integer $N$~\cite{KT,PGH}).

Any finite field lifts the degeneracy of the ground state. QPT then turns 
to a \textit{crossover}, and the sharp step in the dependence of $M(N_0)$ 
is smeared. The crossover takes place in a narrow interval of gate voltages
$|N_0^\past\!-N_0^\ast|\lesssim\Delta_M$. We expect that at a sufficiently 
low field the crossover width $\Delta_M$ remains to be small compared to 
$\Delta_N$, see Fig.~\ref{fig1}(b). In order to estimate $\Delta_M$, we 
now construct an effective Hamiltonian $H_\text{QPT}$ for the vicinity 
of the transition. 

Such Hamiltonian should be applicable at low energies 
($B,T\ll \Gamma$) and for $N_0$ in the range 
$|N_0^\past\!-N_0^\ast|\ll\Delta_N$, which includes the crossover 
region. At these energies and gate voltages the number of electrons 
in the dot is approximately constant, $N\approx N^*$, while $M(N_0)$ 
changes rapidly. It is therefore plausible that $H_\text{QPT}$ acts 
only on the spin degrees of freedom (spin-charge separation). At 
energies below $\Gamma$ half of the dot's spin when it is in the 
triplet state is already screened. The simplest possible model 
accounting for the interaction of the (still unscreened) spin-$1/2$ 
with electrons in the narrow strip of energies $|\xi_k|\lesssim\Gamma$ 
reads
\beq
H_\text{QPT} = H_0^\prime 
+ J^\prime\!\left({\bf s}^\prime\!\cdot\!{\bf S}\right) - B S^z.
\label{19}
\eeq
Here $H_0^\prime$ and ${\bf s}^\prime$ [cf. $H_0$ and ${\bf s}$ in \Eq{10}]  
are defined in terms of the operators $\psi_{ks}^\prime$ acting in the  
basis of single-particle states that incorporate an extra scattering 
phase shift $\delta^* = \pi N^*\!/2$. 

For $H_\text{QPT}$ to describe the change of the ground 
state symmetry at $N_0^\past\!=N_0^\ast$, the exchange constant 
$J'$ must change its sign at this point~\cite{PWA,Nozieres,Bethe}.  
Assuming the dependence $J^\prime(N_0)$ to be analytical,
we can write
\beq
\nu J^\prime(N_0)\sim \Delta_N^{-1}\bigl(N_0^\ast\!-N_0^\past\!\bigr).
\label{20}
\eeq
The coefficient in \Eq{20} has been chosen in such a way that
$\nu J^\prime\sim 1$ for $N_0^\ast\!-N_0^\past\sim\Delta_N$. 
This ensures the continuity of $M(N_0)$ throughout the 
singlet side of the transition $N_0^\past\!<N_0^\ast$. 

A comment on the status of Eqs. \eq{19}, \eq{20} is in order here.
The effective low-energy Hamiltonian $H_\text{QPT}$ is in the same 
relation to the original microscopic model \eq{4}-\eq{7} as, e.g., 
the effective Fermi-liquid description of strong coupling regime~\cite{Nozieres} 
is to the Kondo model. As in the latter case, the applicability of 
$H_\text{QPT}$ can be verified by comparing the predictions of 
the two models.

The magnetization for the model \eq{19}, \eq{20} is obtained using 
the standard scaling arguments~\cite{PWA}.
Very close to the transition (when $|\nu J'|\ln(\Gamma/B)\ll 1$) the first
order perturbation theory in $\nu J'\ll 1$ yields
\beq
M -1 \approx -\nu J'\!/2\sim \Delta_N^{-1}\bigl(N_0^\past\!-N_0^\ast\bigr).
\label{21} 
\eeq
On the doublet side $M(N_0)$
slowly increases with the distance to the transition, saturating at
\beq
M = 1 + \bigl[2\ln(\Gamma/B)\bigr]^{-1}.
\label{22} 
\eeq
Note that \Eq{22} matches \Eq{17} at the border of the 
mixed-valence region.
 
On the singlet side of the transition $(N_0^\past\!<N_0^\ast)$ 
the magnetization is given by \Eq{15} (with $M_0=1$) for 
$B\gg T_K$ and by \Eq{16} for $B\ll T_K$, where the Kondo 
temperature $T_K(N_0)$ satisfies 
\beq
\ln(\Gamma/T_K) = (\nu J^\prime)^{-1}
\sim\Delta_N\bigl(N_0^\ast\!-N_0^\past\!\bigr)^{-1}.
\label{23}
\eeq
$T_K$ increases with the distance to the QPT from $T_K=0$ at 
$N_0^\past\!=N_0^\ast$ to $T_K\!\sim\Gamma$ at the border 
of the mixed valence region, where it matches \Eq{14}. 

As $N_0$ is tuned through the mixed-valence region
$M$ grows monotonically from $M\!\sim B/\Gamma\ll 1$ 
to the value given by \Eq{22}.
The increase takes place mainly in a narrow interval on the singlet 
side of the transition where $B\lesssim T_K(N_0)$.
\Eq{23} then yields the estimate 
\beq
\Delta_M\sim \frac{\Delta_N}{\ln(\Gamma/B)}\,.
\label{24}
\eeq

The evolution of the phase shifts with $N_0$ can now be deduced 
from \Eq{8}. To the left of the crossover [see Fig.~\ref{fig1}(b)] 
both phase shifts are given by $\delta_s\approx \pi N^\ast/2$
with $N^*\!\approx \widetilde{N}_0$, see \Eq{9}. As $N_0$ 
is tuned through the crossover, $\delta_\uparrow$ raises, while 
$\delta_\downarrow$ drops by approximately $\pi/2$. Therefore 
the phase shifts necessarily pass through, respectively, the 
anti-resonance $\delta_\uparrow = 0\,(\text{mod}\,\pi)$
and resonance $\delta_\downarrow= \pi/2\,(\text{mod}\,\pi)$.
Hence, within the crossover region the conductances \eq{3} 
satisfy $g_\uparrow/g_\downarrow\ll 1$, and $g_\uparrow$ 
vanishes identically at some value of $N_0$. In other words, 
the system acts as a \textit{perfect spin filter}. 

Details of the dependencies $g_s(N_0)$ are sensitive to 
the dot's occupation at the transition $N^*$. While $N^*$ 
is close to a half-integer, it's precise value is obviously 
non-universal. For example, $N^*$ depends on the values 
of $t_n$ for all $n$ in \Eq{6}. In Fig. \ref{fig1}(c) we sketch 
$g_s(N_0)$ for $0<\alpha\ll 1$, where $\alpha = N^\ast\!-\widetilde{N}_0$. 
The dependence of the total conductance $g$ on $N_0$ in this 
case has a characteristic Fano-like shape, see Fig. \ref{fig1}(d). 

\begin{figure}[h]
\includegraphics[width=0.9\columnwidth]{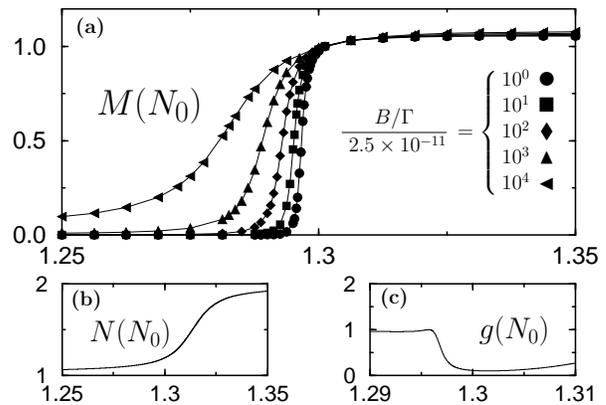}
\caption{
Results of NRG simulations of the model \eq{4}-\eq{7},\eq{25} 
with $\Gamma/E_C =0.05$,  $E_S/E_C = 0.16$, and $\varepsilon/E_C = 0.1$.
\\
(a) Magnetization $M(N_0)$ at different $B$. At the transition $M=1$ 
independently of $B$, in agreement with \Eq{21}.  
(b) Dot's occupation $N(N_0)$. 
(c) Conductance $g(N_0)$ at $B/\Gamma=2.5\times 10^{-11}$. 
Note that in this case $\alpha = N^\ast\!-3/2\approx-0.25<0$,
hence the difference with Fig.~\ref{fig1}(d).
}
\label{fig3}
\end{figure} 

In order to verify the applicability of the effective Hamiltonian 
\eq{19}, we performed extensive numerical renormalization 
group (NRG)~\cite{Wilson} simulations. For this purpose, 
we truncated the dot's Hamiltonian \eq{7} to that of a two-level 
system~\cite{PGH} with
\beq
\epsilon_n = n\varepsilon/2,
\quad 
n=\pm 1.
\label{25}
\eeq
The NRG data, see Fig.~\ref{fig3}, are indeed in an excellent 
agreement with the behavior expected from \Eq{19}.
The sharpening of the step in the dependence $M(N_0)$ with the 
decrease of $B$, obvious in Fig.~\ref{fig3}(a), is described very 
well by
$
\Delta_M\!/\Delta_N = a\bigl[\ln^{-1}(\Gamma/B)+ \,b \ln^{-2}(\Gamma/B)\bigr] 
$
with $a=3.0$ and $b=9.5$; at low field this agrees with \Eq{24}. 
Here we  defined $\Delta_M$ as the distance in $N_0$ between the 
points where $M=0.5$ and $1$, and $\Delta_N$ as the distance between 
the points in Fig.~\ref{fig3}(b) where $N=1.25$ and $1.75$. 

So far, we considered the conductance at $T=0$. The above results 
are valid as long as $T\ll B$; corrections to $g_s$ in this case are of 
the order of $(T/B)^2$, and the spin-filtering property remains intact:
$\min\{g_\uparrow/g_\downarrow\}\sim(T/B)^2\ll 1$. At $T\gg B$ the field 
has a negligible effect. The dependence $g(N_0)$ in this limit is very similar 
to that at $B\gg T$, see Fig.~\ref{fig1}(d) and~\ref{fig3}(c), with $T$ 
replacing $B$ in the crossover width \Eq{24}. This peculiar dependence 
will be observable already at moderately low temperatures $T\lesssim\Gamma$ 
(note that the observability of the conventional Kondo effect requires 
$T\lesssim \min\{T_K\}\ll\Gamma$).
 
To conclude, we studied a lateral quantum dot asymmetrically 
coupled to two conducting leads, and tuned to the mixed-valence 
region between the Coulomb blockade valleys with $S=1/2$ 
and $S=1$ on the dot. This regime can be realized in devices such as
those studied in \cite{two-stage}.
We predict that, contrary to naive expectations, the conductance varies 
with the gate voltage on the scale which is parametrically small 
compared with the width of the mixed-valence region. 

\begin{acknowledgments}
We thank N. Andrei, V. Cheianov, V.I. Falko, L.I. Glazman, 
A.J. Millis, and S. Tarucha for valuable discussions. 
This work was supported by the Nanoscience/ Nanoengineering 
Research Program of Georgia Tech, 
by EC RTN2-2001-00440 ``Spintronics'', Project OTKA D048665, 
and by the J\'anos Bolyai Scholarship.
\end{acknowledgments}


\begin{thebibliography}{99}
\bibitem{spin_1}
A dot with even $N$ will have $S=1$ if the spacing between the 
two single-particle levels closest to the Fermi level is very 
small~\cite{review,spin_theory,spin_exp,real}. This condition 
is easy to satisfy since the spacing is tunable~\cite{two-stage}.  
However, having $S>1$ (e.g., $S=3/2$ for odd $N$)
requires three or more levels to be nearly degenerate, 
which happens rarely, if ever.
\bibitem{phase_shifts}
Note that in \Eq{3} the phase shifts are defined $\text{mod}\,\pi$ 
(i.e. $\delta_s$ and $\delta_s+\pi$ are equivalent). The ambiguity 
is removed in \Eq{8} by setting $\delta_s=0$ for $N_0\to-\infty$.
\end{thebibliography}
\end{document}